          \documentclass{cmslatex}
\usepackage[paperwidth=7in, paperheight=10in, margin=.875in]{geometry}

\usepackage{amssymb,mathrsfs}
\usepackage{amsmath}
\usepackage{graphicx}
\usepackage{dcolumn}
\usepackage{bm}
\usepackage{color}
\usepackage{epsfig}
\usepackage{epstopdf}
\usepackage{float}
\usepackage{textcomp}

\renewcommand{\theequation}{\arabic{section}.\arabic{equation}}

\newtheorem{thm}{Theorem}[section]

\newtheorem{rem}[thm]{Remark}

\allowdisplaybreaks

\begin{document}
\title{Non-Isothermal Electrokinetics: Energetic Variational Approach}

\author{Pei Liu\thanks{Department of Mathematics, Pennsylvania State University, University Park, PA 16802, USA (pul21@psu.edu).}\and{Simo Wu\thanks{Department of Mathematics, Pennsylvania State University, University Park, PA 16802, USA (szw184@psu.edu).}}\and{Chun Liu \thanks{Department of Applied Mathematics, Illinois Institute of Technology, Chicago, IL 60616, USA (cliu124@iit.edu).}}}

\thanks{The research is partially supported by NSF grants DMS-1714401, DMS-1412005.  The authors would like to thank Prof. Zhenli Xu from Shanghai Jiao Tong University for the valuable discussion and constructive comments. The authors also thank the great working environment and support from the Department of Mathematics, Pennsylvania State University. }




\pagestyle{myheadings} \markboth{Non-Isothermal Electrokinetics}{P. Liu, S. Wu, C. Liu} \maketitle

\begin{abstract}
Fluid dynamics accompanies with the entropy production thus increases the local temperature, which plays an important role in charged systems such as the ion channel in biological environment and electrodiffusion in capacitors/batteries. In this article, we propose a general framework to derive the transport equations with heat flow through the Energetic Variational Approach. According to the first law of thermodynamics, the total energy is conserved and we can use the Least Action Principle to derive the conservative forces. From the second law of thermodynamics, the entropy increases and the dissipative forces can be computed through the Maximum Dissipation Principle. Combining these two laws, we then conclude with the force balance equations and a temperature equation. To emphasis, our method provide a self consistent procedure to obtain the dynamical equations satisfying proper energy laws and it not only works for the charge systems but also for general systems.
\end{abstract}


\section{Introduction.} The inhomogeneous and time-dependent temperature could be of great importance in the electrodiffusion processes. It also plays a key role in many biological and chemical applications. For example, a number of ion channels are observed to be sensitive to the temperature changes  \cite{Cesare1999,Reubish2009}. These temperature-gated ion-channels can detect the temperature thus regulate the  internal homeostasis and disease-related processes such as the thermal adaptation and the fever response. Also the electro-osmotic flow (EOF) in the microfluidic devices will cause the internal heat generation, which is known to be the Joule heating effects \cite{Grushka1989,Knox1994}. This inhomogeneous increase of the temperature will change the fluid dynamical properties, thus it is important in controlling the EOF and designing microfluidic devices.

The ionic transport can be modeled through the classical Poisson--Nernst--Planck (PNP) theory and its various modified versions \cite{Qiao2014,Im2002,Eisenberg2007,Gillespie2002,Liu2014,Wei2012,XML:PRE:2014,Hsieh2015}, which are shown to be successful in describing various phenomenon and properties. Through the energetic variational approach (EnVarA), C. Liu et al. derived the modified PNP equations with given free energy functional and the form of entropy production \cite{HTLE:JPC:12,HEL:CMS:11,SXu2014}. However, these models are all isothermal: the temperature is fixed as a constant.

To model the non-isothermal dynamic processes, we need to couple the mechanical equation and the thermal equation together. In \cite{Feireisl2007}, Feireisl considered the incompressible homogeneous Newtonian fluids with temperature dependent coefficients and obtained the long-time and large-data existence for a suitable weak solution. For the heat conducting compressible Newtonian fluid, Bulicek \cite{Bulicek2009} developed a Navier--Stokes--Fourier system and derived priori estimates and the weak stability based on variational weak formulation and the thermodynamic second law. In \cite{Eleuteri2015}, Eleuteri studied the non-isothermal diffuse-interface model for two incompressible Newtonian fluids, resulting with a Cahn-Hilliard system. Also, many papers are devoted into real applications. For example, Xuan et al. \cite{XuanLi2008} reported the Joule heating effects in the electrokinetic flow could increase the current load, enhance the flow rate and reduce the separation efficiency. S\'anchez et al. \cite{Sanchez2013} analyzed the Joule heating effect on a purely electroosmotic flow of non-Newtonian fluids through a slit microchannel . In addtion, Gonzalez et al. \cite{GONZALEZ2006} studied the electrothermal motion in microsystems generated by AC electrical field. 

In this work, we adapt the EnVarA, aiming to propose an unified framework to self-consistently describe the electrothermal motion. With given form of the free energy functional and the entropy production, the conservative forces can be derived through the Least Action Principle (LAP) and the dissipative forces are given by the Maximum Dissipation Principle (MDP). For any open subset of the fluid region, the energy balance and the entropy increase lead to the mechanical and thermal equations. Here we derive the model in Euler coordinates, in contrast to the classical approach in literature based on the Lagrange formalism. The reason is the charged systems usually involve more than one ionic species, thus several velocity fields appear, employing material derivative like previous papers might cause confusion and inconvenience.

To emphasize, our approach guarantees the resulting equations satisfying fundamental laws of thermodynamics and it can be generalized to a variaty of complex fluid systems such as liquid crystal \cite{FranChun2017}.

\section{Theory of Charge Dynamics.} Consider a closed system with $N$ ionic species in domain $\Omega$. Define the local density distribution for $i$th species as space dependent: $\rho_i(\mathbf{r},t),~ i=1,2,\cdots,N,$ and the valences are denoted by $z_i$ respectively. The time evolution is usually modeled through the PNP equation:
\begin{equation}
\begin{cases}
\displaystyle   \frac{\partial}{\partial t} \rho_i = \nabla \cdot D_i \left( k_B T \nabla \rho_i + \rho_i z_i e \nabla \phi \right),\\
 -\nabla \cdot \epsilon \nabla \phi =\sum_{i=1}^N \rho_i z_i e.
\end{cases}
\label{PNP_equation}
\end{equation} 
Here $\phi$ is the mean electrical potential, $e$ is the elementary charge, $D_i$ is the diffusion coefficient, $k_B$ is the Boltzmann constant. And $T$ is the temperature, which is homogeneous and remains independent of time within the PNP model. The corresponding free energy dissipation is known to be,
\begin{equation}
\frac{d}{dt} \int_\Omega \left(\frac{\epsilon}{2} |\nabla \phi|^2  + \sum_{i=1}^N  k_B T \rho_i (\log \rho_i -1) \right)d\mathbf{r} = -\sum_{i=1}^N \int_\Omega \frac{D_i}{\rho_i} |k_B T \nabla \rho_i + \rho_i z_i e \nabla \phi|^2 d\mathbf{r},
\label{PNP_energy_law}
\end{equation}
where the left integral is the total free energy including the mean electrical energy and the entropic contribution. And the right hand side represents for the energy dissipation. With the energy dissipation law \eqref{PNP_energy_law}, the PNP equation \eqref{PNP_equation} can be derived through the EnVarA \cite{SXu2014}. In the PNP theory, the system entropy increases without affecting the temperature, which indicates a specific amount of heat/energy must be transferred to the system. Through the second law of thermodynamics, it is straightforward to evaluate the heat absorption rate $\frac{dQ}{dt}=-\sum_{i=1}^N \int_\Omega \frac{D_i}{\rho_i} (k_B T \nabla \rho_i + \rho_i z_i e \nabla \phi) \cdot \rho_i z_i e \nabla \phi\ d\mathbf{r}$. When the heat conducting rate is very large or the total heat generated is negligible compared with the system heat capacitance, it is reasonable to assume the temperature $T$ is a constant. But more generally, we have to consider the temperature evolution with given heat sources.

\subsection{Energy functional.} To study the effects of temperature, we use $T(\mathbf{r},t)$ to describe the temperature distribution at time $t$. We write the general form of the free energy $F(V,t)$ for the system in any subdomain $V \subset \Omega$ which is a functional of the temperature and particle densities,
\begin{eqnarray}
F(V,t)&=& \sum_{i=0}^N \int_\Omega \Psi_i(\rho_i(\mathbf{r},t),T(\mathbf{r},t)) d\mathbf{r} + \sum_{i,m=0}^N\frac{z_i z_m e^2}{2} \iint_V \rho_i(\mathbf{r},t) \rho_m(\mathbf{r'},t) v(\mathbf{r},\mathbf{r'}) d\mathbf{r}d \mathbf{r'}\nonumber\\
&&+ \sum_{i=0}^N z_i e \int_V \rho_i(\mathbf{r}) \left(\psi(\mathbf{r},t) + \sum_{m=1}^N z_m e \int_{\Omega \backslash V} \rho_m(\mathbf{r'},t) v(\mathbf{r},\mathbf{r'}) d\mathbf{r'}\right)d \mathbf{r}.
\label{FreeEnergy_g}
\end{eqnarray}
The first term $\Psi_i$ is a local function of density $\rho_i(\mathbf{r},t)$ and temperature $T(\mathbf{r},t)$, representing the free energy density from the entropy contribution. The index $i=0$ stands for the solvent particles, which is imcompressible with constant density $\rho_0$, and index $1,\cdots,N$ represents the solute species. The second term in \eqref{FreeEnergy_g} represents for the potential energy  from the Coulomb interaction $z_i z_m e^2v(\mathbf{r},\mathbf{r}')$. The last term is the potential energy from the external field, including the external electrical potential $\psi$ and the contribution from particles outside domain $V$. Since the negative local entropic density is the derivative of the free energy density with respect to the temperature, the entropy,
\begin{equation}
S(V,t)= - \sum_{i=0}^N \int_V  \frac{\partial \Psi_i(\rho_i(\mathbf{r},t),T(\mathbf{r},t))}{\partial T(\mathbf{r},t)} d\mathbf{r}.
\label{charge_entropy_functional}
\end{equation}
Then the corresponding internal energy is given by the Legendre transform,
\begin{eqnarray}
U(V,t) &=&  \sum_{i=0}^N \int_V \left(\Psi_i(\rho_i(\mathbf{r},t),T(\mathbf{r},t))-T(\mathbf{r},t)\frac{\partial \Psi_i(\rho_i(\mathbf{r},t),T(\mathbf{r},t))}{\partial T(\mathbf{r},t)}\right) d\mathbf{r} \nonumber \\
&&+\sum_{i,m=0}^N \frac{z_i z_m e^2}{2}\iint_V \rho_i(\mathbf{r},t) \rho_m(\mathbf{r'},t) v(\mathbf{r},\mathbf{r'})d\mathbf{r} d \mathbf{r'}\nonumber\\
&&+ \sum_{i=0}^N z_i e \int_V \rho_i(\mathbf{r}) \left(\psi(\mathbf{r},t) + \sum_{m=1}^N z_m e \int_{\Omega \backslash V} \rho_m(\mathbf{r'},t) v(\mathbf{r},\mathbf{r'}) d\mathbf{r'}\right)d \mathbf{r}.
\label{internal_functional}
\end{eqnarray}

For each species, the velocity field is denoted as $u_i(\mathbf{r},t)$, so the ionic densities satisfy the conservation law: $\frac{\partial}{\partial t} \rho_i +\nabla\cdot  (\rho_i u_i) =0$. Each velocity field $u_i(\mathbf{x_i}(\mathbf{X},t),t)$ determines an unique flow map  $\mathbf{x_i}(\mathbf{X},t)$ for the corresponding particle species through $\frac{\partial}{\partial t} \mathbf{x_i}(\mathbf{X},t) = u_i(\mathbf{x_i}(\mathbf{X},t),t)$. Introduce the mass $m_i$ of each species and the kinetic energy $ K(V,t)=\frac{1}{2}\sum_{i=0}^N \int_V m_i \rho_i(\mathbf{r},t) u_i^2(\mathbf{r},t) d\mathbf{r}$. Then the total action of the whole system is $A = -\int_0^T \left[K(\Omega,t)+F(\Omega,t)\right] dt$. According to the LAP, the conservative force on each species can be obtained through variation of action with respect to the flow map,
\begin{equation}
f_i^\text{con}(\mathbf{x},t)= \frac{\delta A}{\delta \mathbf{x_i}(\mathbf{X},t)}
=-m_i \rho_i \left(\frac{\partial}{\partial t} u_i + u_i \nabla  u_i\right)-\nabla P_i -\rho_i z_i e \nabla \phi, 
\end{equation}
where the mean electrical potential $\phi(\mathbf{r},t) = \psi(\mathbf{r}) + \sum_{j=0}^N  z_j e \int_{\Omega} \rho_j(\mathbf{r}',t) v(\mathbf{r},\mathbf{r}') d\mathbf{r}'$. And  $  P_i(\mathbf{r},t) = \rho_i^2(\mathbf{r},t)\frac{\partial}{\partial \rho_i(\mathbf{r},t)}\left(\frac{\Psi_i(\rho_i(\mathbf{r},t),T(\mathbf{r},t))}{\rho_i(\mathbf{r},t)}\right)=\rho_i(\mathbf{r},t) \frac{\partial \Psi_i(\rho_i(\mathbf{r},t),T(\mathbf{r},t))}{\partial \rho_i(\mathbf{r},t)} -\Psi_i(\rho_i(\mathbf{r},t),T(\mathbf{r},t)),$ is the thermal pressure \cite{SXu2014} for $i=1,\cdots,N$. Since the solvent is incompressible, the thermal pressure $P_0$ appears as a Lagrange multiplier and is undetermined. 

\newpage

\begin{rem}
We can define the entropic density $\eta_i(\mathbf{r},t) =- \frac{\partial \Psi_i(\rho_i(\mathbf{r},t),T(\mathbf{r},t))}{\partial T(\mathbf{r},t)} $ and internal energy density $e_i^{\text{int}}(\rho_i(\mathbf{r},t),\eta_i(\mathbf{r},t)) = \Psi_i(\rho_i(\mathbf{r},t),T(\mathbf{r},t))+T(\mathbf{r},t)\eta_i(\mathbf{r},t)$, then the definition of thermal pressure is equivalent to $P_i(\mathbf{r},t) = \rho_i^2(\mathbf{r},t)\frac{\partial}{\partial \rho_i(\mathbf{r},t)}\left(\frac{e_i^{\text{int}}(\rho_i(\mathbf{r},t),\eta_i(\mathbf{r},t))}{\rho_i(\mathbf{r},t)}\right)$.
\end{rem}

According to the first law of thermodynamics, the internal energy is conserved with the work done and the heat absorbed. The rate of work is given by,
\begin{eqnarray}
\frac{d}{dt} W(V,t) &=&\sum_{i=0}^N z_i e\int_{V}  \rho_i(\mathbf{r},t)  \frac{\partial }{\partial t} \left[\psi(\mathbf{r},t) +\sum_{m=0}^N z_m e\int_{\Omega \backslash V} \rho_m(\mathbf{r'},t)v(\mathbf{r},\mathbf{r'}) d\mathbf{r'}\right]  d\mathbf{r} \nonumber\\
&&+\sum_{i=0}^N \int_{\partial V} \mathbb{T}_i(\mathbf{r},t) u_i(\mathbf{r},t) \cdot  d\mathbf{r}.
\label{Wt}
\end{eqnarray}
Here the first integral is due to the time dependent external field, including the contribution from ions in domain $\Omega \backslash V$. The second term is from the work of the stress tensor $\mathbb{T}_i$ on the boundary $\partial V$, which includes the contribution from the thermal pressure $P$ and the dissipative force. The form of $\mathbb{T}_i(\mathbf{r},t)$ will be specified when we have the dissipative force in Eq. \eqref{dis_force}. The rate of heat transfer,
\begin{equation}
\frac{d}{dt} Q(V,t) = -\int_{\partial V}  j \cdot d\mathbf{r}+\int_{V} q d\mathbf{r}.
\label{Qt}
\end{equation}
We should notice here the control volume $V$ does not move along with the velocity field as different solute species have different flow maps. At the boundary $\partial V$, the mechanical flux will also introduce an total energy flux, which should be considered,
\begin{equation}
J_E(V,t) = -\sum_{i=0}^N \int_{\partial V} u_i \left[\frac{1}{2}m_i\rho_i u_i^2+e_i^{\text{int}}(\rho_i,\eta_i)+z_i e\rho_i\phi \right]d \mathbf{r}.
\label{boundary_internalenergy}
\end{equation}
So, the energy conservation is expressed as,
\begin{equation}
\frac{d}{dt}\left[U(V,t)+K(V,t)\right] = \frac{d}{dt} W(V,t) + \frac{d}{dt} Q(V,t) + J_E(V,t).
\label{firstlaw}
\end{equation}
With \eqref{internal_functional} and \eqref{Wt}--\eqref{firstlaw}, and use the fact that the control volume $V$ is arbitrary chosen, we then obtain a differential equation,
\begin{eqnarray}
&&\sum_{i=0}^N \left(- T \frac{\partial^2 \Psi_i}{\partial T^2}\right) \left(\frac{\partial T}{\partial t} + u_i \cdot \nabla T\right) +\sum_{i=1}^N \left(\Psi_i - \rho_i \frac{\partial \Psi_i}{\partial \rho_i}-T\frac{\partial \Psi_i}{\partial T}+\rho_i T\frac{\partial^2 \Psi_i}{\partial T\partial \rho_i}\right) \nabla \cdot u_i \nonumber \\
&  +&\sum_{i=0}^N \left[\rho_i u_i  z_i e\nabla \phi -\nabla \cdot\left(\mathbb{T}_i u_i \right)\right]+ \sum_{i=0}^N m_i \rho_i (\frac{\partial}{\partial t} u_i + u_i \nabla u_i)\cdot u_i  =q-\nabla \cdot j.
\label{conserve_internal}
\end{eqnarray}
This equation provides a relation between the temperature evolution and the mechanical velocities. In order to form a closed PDE system, we need another relation which can be given by the entropy production.

\subsection{Entropy production.} The second law of thermodynamics states the fact that, the entropy increase of any closed system must not be less than the heat absorbed from the environment, and the equality only holds for reversible process. We choose the entropy production rate of the system in arbitrary domain $V$ to be,
\begin{equation}
\Delta(V,t) = \int_{V} \left(\sum_{i=1}^N \frac{\nu_i \rho_i |u_i-u_0|^2 + \xi_i |\nabla \cdot u_i|^2}{T} +\sum_{i=0}^N\frac{ \lambda_i |\nabla u_i|^2}{T} + \frac{1}{k} |\frac{j}{T}|^2 \right)d\mathbf{r}.
\label{EP}
\end{equation}
where  $\xi_i$ is the bulk viscosity, $\lambda_i$ is the shear viscosity coefficient for $i$th species. $\nu_{i}$ describes the viscosity between the $i$th particle and the solvent. And $j$ represents for the heat flux, $k$ is a constant relating with the heat conductance. Here we only consider the relative drag between solvent and solute while neglecting the friction between different solvent species. This is generally true for dilute solutions, and the correction can be made following the argument in \cite{Hsieh2015}. Compared with the entropy production in the classical fluid dynamic equations, we have one extra term from the heat flux. Then the dissipative force is given by the MDP,
\begin{equation}
f^{\text{dis}}_{i}(\mathbf{r},t) = \frac{T}{2}\frac{\delta \Delta(\Omega,t)}{\delta u_i(\mathbf{r},t)}=\begin{cases}  \nu_i \rho_i (u_i-u_0)-\nabla (\xi_i \nabla \cdot u_i) - \nabla \cdot \lambda_i \nabla u_i,\hspace{5pt} i=1,\cdots,N.\\
\sum_{m=1}^N \nu_m \rho_m(u_0 - u_m)- \nabla \cdot \lambda_0 \nabla u_0, \hspace{17pt} i=0. 
\end{cases}
\label{dis_force}
\end{equation}
So the stress tensor $\mathbb{T}_i = [-P_i+ (\xi_i-\lambda_i) \nabla \cdot u_i]\mathbb{I} + 2\lambda_i \mathcal{D} u_i$, for $i=1,\cdots,N$, where $\mathbb{I}$ is an 3-by-3 identity matrix, $\mathcal{D} u_i=[\nabla u_i + (\nabla u_i)^T]/2 $ is the symmetric part of $\nabla u_i$. For the imcompressible solvent, $\mathbb{T}_0 = -P_0\mathbb{I} + 2\lambda_0 \mathcal{D} u_0$. Similarly, we should take into account the entropic flux at boundary $\partial V$,
\begin{equation}
J_S =  \sum_{i=0}^N \int_{\partial V} \frac{\partial \Psi_i(\rho_i(\mathbf{r},t),T(\mathbf{r},t))}{\partial T(\mathbf{r},t)} u_i \cdot d\mathbf{r}.
\label{JS}
\end{equation}
So the second law of thermodynamics is expressed as,
\begin{equation}
\frac{d }{dt}S(V,t) + \int_{\partial V} \frac{j}{T} \cdot d\mathbf{r} - \int_{V} \frac{q}{T}  d\mathbf{r} -J_S=\Delta(V,t)\geq 0,
\label{EP_def}
\end{equation}
where $q$ is the heat source. Combining \eqref{charge_entropy_functional}, \eqref{EP} and \eqref{EP_def}, we then obtain,
\begin{equation}
-\sum_{i=0}^N \left[ \frac{\partial^2 \Psi_i}{\partial T^2}\left(\frac{\partial T}{\partial t}+u_i \cdot \nabla T  \right)+ \left(\frac{\partial \Psi_i}{\partial T}-\rho_i\frac{\partial^2 \Psi_i}{\partial T \partial \rho_i}\right) \nabla \cdot u_i \right] =\widetilde{\Delta} +\frac{q}{T}-\nabla \cdot \frac{j}{T}.
\label{increase_entropy}
\end{equation}
Here $\widetilde{\Delta}$ is the entropy production density, $\Delta(V,t)=\int_V \widetilde{\Delta}(\mathbf{r},t)d\mathbf{r}$. 

\subsection{Governing equations.} Combining \eqref{conserve_internal} and \eqref{increase_entropy} gives, 
\begin{equation}
\sum_{i=0}^N u_i \cdot \left( f_i^{\text{con}} - f_i^{\text{dis}}\right) = j\cdot \left(\frac{j}{kT}+ \frac{\nabla T}{T}\right).
\label{balance}
\end{equation}
The right hand side is about the heat flux which is invariant under any inertial frames of reference, while the left hand side is about the particle velocity which depends on the reference frame we choose. So it is reasonable to claim the coefficients must vanish, thus the Onsager Principle holds, $f_i^{\text{con}} = f_i^{\text{dis}}$, for $i=0,\cdots,N$. And the heat flux: $j=-k\nabla T$, which is the Fourier law. Then the dynamic equations for the solute particles become,
\begin{equation}
\begin{cases}
\displaystyle \frac{\partial}{\partial t} \rho_i + \nabla \cdot (\rho_i u_i)=0 ,\\
\displaystyle m_i \rho_i (\frac{\partial u_i}{\partial t} + u_i \nabla  u_i)+\nabla P_i+\rho_i z_ie\nabla \phi=\nu_i \rho_i (u_0-u_i)+\nabla (\xi_i \nabla \cdot u_i) + \nabla \cdot \lambda_i \nabla u_i,\\
-\nabla \cdot \epsilon \nabla \phi =\sum_{m=1}^N \rho_m z_m e+\rho_f.
\end{cases}\hspace{-0.5cm}
\label{gPNP_solute}
\end{equation} 
Here we use the fact that $-\nabla \cdot \epsilon \nabla v(\mathbf{r},\mathbf{r'}) = \delta(\mathbf{r}-\mathbf{r'})$, where $\epsilon$ is the dielectric constant. And $\rho_f = -\nabla \cdot \epsilon \nabla \psi$ describes the external field. For the imcompressible solvent,
\begin{equation}
\begin{cases}
\displaystyle  m_0 \rho_0 \left(\frac{\partial}{\partial t} u_0 + u_0 \nabla  u_0\right)+\nabla P_0 +\rho_0 \nabla \phi_0=\sum_{i=1}^N\nu_i \rho_i (u_i-u_0) + \nabla \cdot \lambda_0 \nabla u_0,\\
  \nabla \cdot u_0 = 0.
\end{cases}
\label{gPNP_solvent}
\end{equation}

\newpage

And the temperature equation,
\begin{align}
&\sum_{i=0}^N \left(- T \frac{\partial^2 \Psi_i}{\partial T^2}\right)\left( \frac{\partial T}{\partial t} +  u_i  \cdot \nabla T\right) + \left(\sum_{i=1}^N  \frac{\partial P_i}{\partial T}\nabla \cdot u_i \right)T  \nonumber\\=& \nabla \cdot k \nabla T +  \sum_{i=1}^N \nu_i \rho_i |u_i-u_0|^2 + \xi_i |\nabla \cdot u_i|^2 +\sum_{i=0}^N \lambda_i |\nabla u_i|^2 +q.
\label{gPNP_temp}
\end{align}
In Eq. \eqref{gPNP_temp}, $ \sum_{i=0}^N \left(- T \frac{\partial^2 \Psi_i}{\partial T^2}\right)$ can be viewed as the weighted averaged heat capacitance of the system. The second term represents the work of thermopressure transfer into heat. On the right hand side, $ \nabla \cdot k \nabla T$ describes the heat diffusion. We should notice the entropy production from mechanical viscosity appears as an internal heat source.

Together with proper boundary conditions, the equations \eqref{gPNP_solute}, \eqref{gPNP_solvent} and \eqref{gPNP_temp} form a closed PDE system to describe the non-isothermal electro-thermal flow. 

\begin{rem}
For real physical system, we usually have $\frac{\partial^2 \Psi_i}{\partial T^2}<0$, corresponding to a positive heat capacitance. For example, $\Psi = k_B T \rho (\log \rho - C\log T)$ and $\frac{\partial^2 \Psi}{\partial T^2}= -\frac{C\rho}{T}$ for ideal gas. Thus $\frac{\partial T}{\partial t}$ and $\nabla \cdot k\nabla T$ in \eqref{gPNP_temp} have the same sign with the heat equation.
\end{rem}

\begin{rem}
The heat source term $\sum_{i=1}^N \nu_i\rho_i |u_i-u_0|^2$ in Eq. \eqref{gPNP_temp} takes the form of ionic flux square times the resistance. If the temperature and densities are all homogeneous in space, then the electrical current as well as the ionic fluxes $\rho_i (u_i-u_0)$ are proportional to the local electrical field. Thus the energy dissipation is equivalent to the well known Joule heating effect, which states the current square times the resistance becomes heat in a circuit. For more general situation, our model suggests to use fluxes of each ionic species instead of the total electrical current.
\end{rem}

\section{Examples} With given free energy density $\Psi_i$ and entropy production $\Delta$, the above approach can be applied and generalized to a wide variety of systems, such as the modified PNP equations with ionic correlation, size effects and relative drags. In this section, we consider two examples.

\subsection{Imcompressible Navier--Stokes--Fourier system} By setting the number of ionic species to be zero, we can also investigate the solvent system alone. Consider the fluid in a confined domain $\Omega$ and at the boundary $\partial \Omega$ there is no mechanical nor heat flux. Since the solvent is imcompressible, the free energy functional is just,
\begin{equation}
F(V,t)=-C \int_V  T(\mathbf{r}) \log T(\mathbf{r}) d\mathbf{r}. 
\end{equation}
Here $C$ is a constant. And the entropy production takes the form,
\begin{equation}
\Delta(V,t) = \int_{V} \left(\frac{  \lambda |\nabla u|^2}{T} + \frac{1}{k} |\frac{j}{T}|^2 \right)d\mathbf{r}.
\end{equation}
Also, the kinetic energy $K(V,t) = \frac{1}{2}\int_V m u^2(\mathbf{r},t) d\mathbf{r}$. So the governing equations become,
\begin{equation}
\begin{cases}
\displaystyle \nabla \cdot u_0 = 0,\\
\displaystyle  m \left(\frac{\partial}{\partial t} u + u \nabla  u\right)+\nabla P = \nabla \cdot \lambda \nabla u,\\
\displaystyle C\left( \frac{\partial T}{\partial t} +  u  \cdot \nabla T\right) = \nabla \cdot k \nabla T + \lambda |\nabla u|^2 +q.
\end{cases}
\end{equation}
This set of equations satisfy the thermodynamic laws automatically,
\begin{equation}
\begin{cases}
\displaystyle \frac{d}{dt}\left[U(\Omega,t)+K(\Omega,t)\right]=\int_{\Omega} q(\mathbf{r},t) d\mathbf{r},\\
\displaystyle \frac{d}{dt}S(\Omega,t) = \int_{\Omega} \left(\frac{q}{T}+\frac{  \lambda |\nabla u|^2}{T} + \frac{1}{k} |\frac{j}{T}|^2 \right)d\mathbf{r},
\end{cases}
\end{equation}
where the entropy is $S(V,t) = C \int_V (\log T(\mathbf{r})+1) d\mathbf{r}$ and the internal energy is given by $U(V,t) = C \int_V T(\mathbf{r}) d\mathbf{r}$. Note the diffusion coefficients $\lambda$ and $k$ can depend on space and the state variables.

\subsection{Poisson--Nernst--Planck--Fourier system} Consider a system in confined domain $\Omega$ and there is no flux at the boundary.  The free energy density function $\Psi_i$ is given by,
\begin{equation}
\Psi_i(\rho_i(\mathbf{r},t),T(\mathbf{r},t)) = k_B T(\mathbf{r},t) \rho_i(\mathbf{r},t) \left[\log \rho_i(\mathbf{r},t) - C_i \log T(\mathbf{r},t)\right],
\label{ideal_free}
\end{equation}
\begin{rem}
In the classical PNP system, the temperature is a constant, so there is only the $\rho \log \rho$ term. Here Eq. \eqref{ideal_free} uses the complete form of ideal gas free energy density as a function of both density and temperature. $C_i$ is a constant related to the heat capacitance of each species.
\end{rem}

And the entropy production rate,
\begin{equation}
\Delta(V,t) = \int_{V} \widetilde{\Delta}(\mathbf{r},t) d\mathbf{r} = \int_{V} \left( \frac{\lambda_0 |\nabla u_0|^2 }{T}+ \sum_{i=1}^N \frac{\nu_i \rho_i |u_i-u_0|^2 }{T} + \frac{1}{k} |\frac{j}{T}|^2 \right)d\mathbf{r}.
\end{equation}
Neglecting the kinetic energy, the governing equations for the ionic species become,
\begin{equation}
\begin{cases}
\displaystyle \frac{\partial}{\partial t} \rho_i + \nabla \cdot (\rho_i u_i)=0 ,\\
\displaystyle \nu_i \rho_i (u_i-u_0) = - k_B \nabla (\rho_i T) - z_i e \rho_i \nabla \phi,\\
-\nabla \cdot \epsilon \nabla \phi =\sum_{m=1}^N \rho_m z_m e+\rho_f.
\end{cases}
\label{mPNP_solute}
\end{equation} 
Here $\rho_f(\mathbf{r},t) = -\nabla \cdot \epsilon \nabla \psi(\mathbf{r},t)$ describes the external field $\psi(\mathbf{r},t)$. And for the solvent,
\begin{equation}
\begin{cases}
\nabla \cdot u_0 = 0,\\
\nabla P_0 + \sum_{i=1}^N \nu_i \rho_i ( u_0 -u_i) - \nabla \cdot \lambda_0 \nabla u_0=0.
\end{cases}
\label{mPNP_solvent}
\end{equation}
And the temperature equation,
\begin{align}
&\left(\sum_{i=0}^N  k_B C_i \rho_i\right) \frac{\partial T}{\partial t} + \left(\sum_{i=0}^N  k_B C_i \rho_i u_i \right) \cdot \nabla T + \left(\sum_{i=1}^N k_B \rho_i  \nabla \cdot u_i\right) T \nonumber\\=& \nabla \cdot k \nabla T +  \lambda_0 |\nabla u_0|^2  + \sum_{i=1}^N \nu_i\rho_i |u_i-u_0|^2 +q.
\label{mPNP_temp}
\end{align}

\begin{rem}
Eq. \eqref{mPNP_solute} is the modified PNP equation, where particles are driven by the pressure gradient and the mean electrical potential. The chemical potential of $i$th ion species is given by, $\mu_i(\mathbf{r},t)= \frac{\delta F(\Omega,t)}{\delta \rho_i(\mathbf{r},t)}= k_B T(\mathbf{r},t) \left[\log \rho_i(\mathbf{r},t) +1 - C_i \log T(\mathbf{r},t)\right]  + z_i e \phi(\mathbf{r},t)$. Thus the equations in \eqref{mPNP_solute} are not equivalent to $\frac{\partial}{\partial t} \rho_i = \nabla \cdot \frac{\rho_i}{\nu_i} \nabla \mu_i$, indicating that when temperature is a variable, we should use pressure instead of using chemical potential. 
\end{rem}

\newpage

\begin{rem}Here we cannot simply assume $u_0$ is a constant, since the solvent energy and entropy are included. This is different from the original PNP equation where the velocity, energy and entropy of the solvent are not considered. Eq. \eqref{mPNP_solvent} might not be solvable without the solvent viscosity $\lambda_0$.
\end{rem}

Eq. \eqref{mPNP_solute}, \eqref{mPNP_solvent}, \eqref{mPNP_temp} form a closed PDE system, which we call Poisson--Nernst--Planck--Fourier (PNPF). We can also check, they satisfy the thermodynamic laws,
\begin{equation}
\begin{cases}
\displaystyle  \frac{d}{dt}U(\Omega,t) = \int_\Omega \left(q+ \sum_{i=0}^N \rho_iz_i e \frac{\partial \psi}{\partial t} \right)d\mathbf{r},\\
\displaystyle   \frac{d}{dt} S(\Omega,t) =  \int_\Omega \left(\frac{q}{T}+ \frac{\lambda_0 |\nabla u_0|^2 }{T}+ \sum_{i=1}^N \frac{\nu_i \rho_i |u_i-u_0|^2 }{T} + \frac{1}{k} |\frac{j}{T}|^2 \right) d\mathbf{r} .
\end{cases}
\label{energy_law_g}
\end{equation}
Note that only for the isothermal system, we can combine the above two thermodynamic laws together, obtaining the free energy dissipation,
\begin{eqnarray}
\frac{d}{dt} F(\Omega,t) &=& \frac{d}{dt} (U(\Omega,t)-TS(\Omega,t)) \nonumber\\
&=& \int_\Omega  \sum_{i=0}^N \rho_i z_i e \frac{\partial \psi}{\partial t} d\mathbf{r}-T\Delta(\Omega,t).
\end{eqnarray}
When the external field is independent of time, this is equivalent to \eqref{PNP_energy_law},

\section{Numerical Results}

In this section, we present the numerical results of the PNPF equations. Consider a one-dimensional channel with $100mM$ $NaCl$ water solution at room temperature $T_0=25 \text{\textdegree}C=298.15 K$, so that $N=2$, $z=\pm 1$ and the water velocity vanishes due to the imcompressible condition. We choose the parameters from the real experimental data. The diffusion coefficients for $Na^+$ is $1.334  nm^2/ns$ and for $Cl^-$ is $2.032 nm^2/ns$ \cite{CRC2004}. Water dielectric permittivity is $78.3$ \cite{malmberg1956dielectric} and 
the specific heat capacity is $75.375J/(mol\cdot K)$ \cite{linstrom2001nist} so that $C_0\rho_0 = 302.15nm^{-3}$. And for an ion, we use $C_\pm = 3$ corresponding to the monomolecular ideal gas. After dimensionless with $\ell=1nm$, $\tau=1ns$, $T_0=298.15K$, the whole set of dimensionless equations becomes,
\begin{equation}
\begin{cases}
\displaystyle \frac{\partial}{\partial t} \rho_i + \nabla \cdot (\rho_i u_i)=0 ,\\
\displaystyle \nu_i \rho_i u_i = - \nabla (\rho_i T) - z_i \rho_i \nabla \phi,\\
\displaystyle -\nabla \cdot \epsilon \nabla \phi =4\pi l_B \sum_{i=\pm} \rho_i z_i,\\
\displaystyle \sum_{i=0,\pm}  C_i \rho_i \frac{\partial T}{\partial t} + \sum_{i=\pm}  C_i \rho_i u_i  \cdot \nabla T + \sum_{i=\pm} \rho_i T \nabla \cdot u_i  = \nabla \cdot k \nabla T +  \sum_{i=\pm} \nu_i\rho_i |u_i|^2.
\end{cases}
\end{equation}
With initial condition $\rho_\pm(x,0)=\rho_0=0.06$ and $T(x,0)=1$. The dimensionless parameters are, $C_0 \rho_0=302$, $C_\pm=3$, $1/\nu_+=1.334$, $1/\nu_-=2.032$, $\epsilon=1$, $l_B=0.714$. In order to highlight the contribution from temperature, we choose a relatively small heat conductance $k=100$. 
The computational domain is $L=10$. The boundary condition for ion density and temperature are Dirichlet, i.e. $\rho_i(0,t)=\rho_i(L,t)=\rho_0$, $T(0,t)=T(L,t)=1$.

\begin{figure}[!htbp]
\begin{center}
\includegraphics[width=0.45\textwidth]{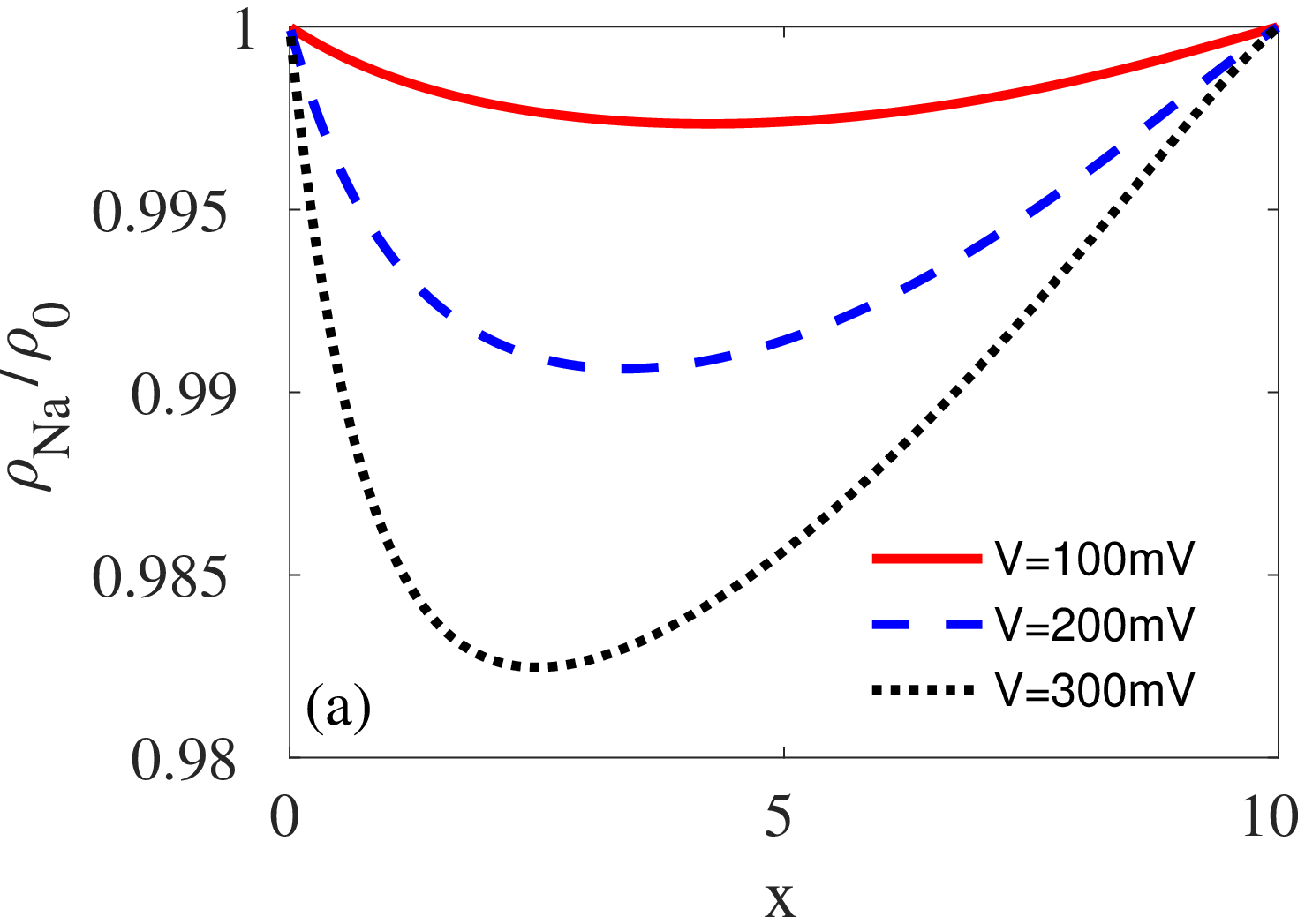}
\includegraphics[width=0.45\textwidth]{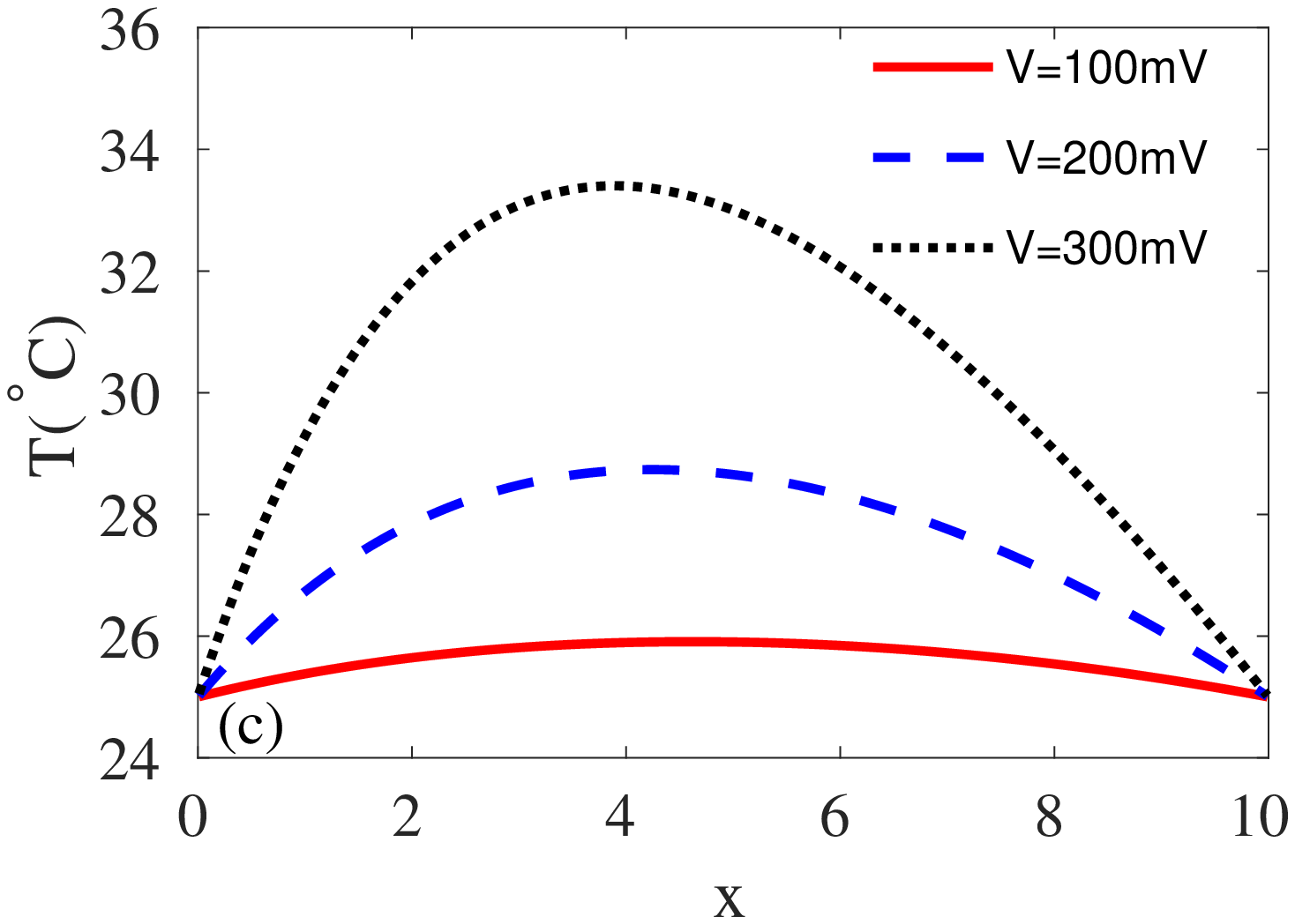}\\
\includegraphics[width=0.45\textwidth]{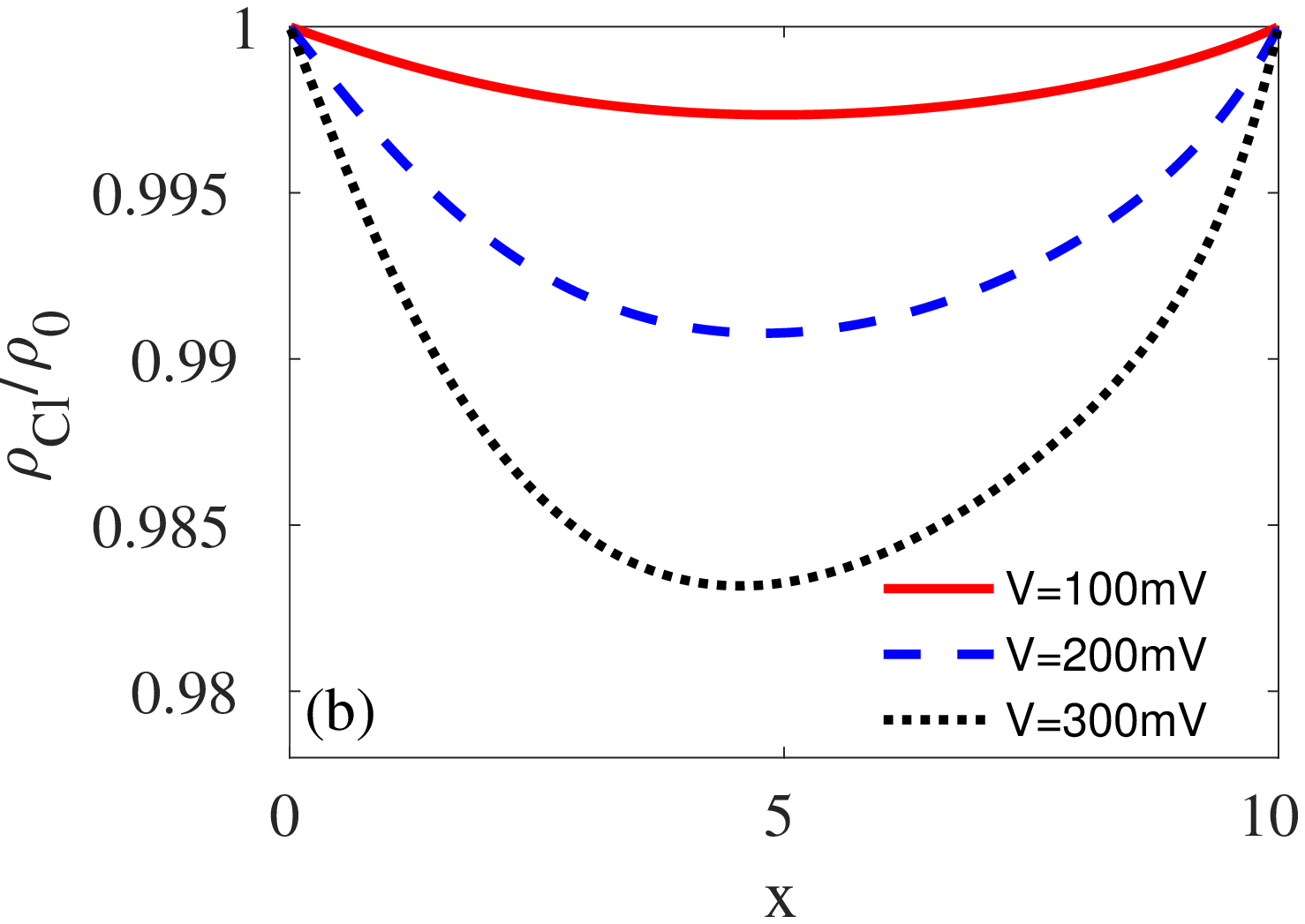}
\includegraphics[width=0.45\textwidth]{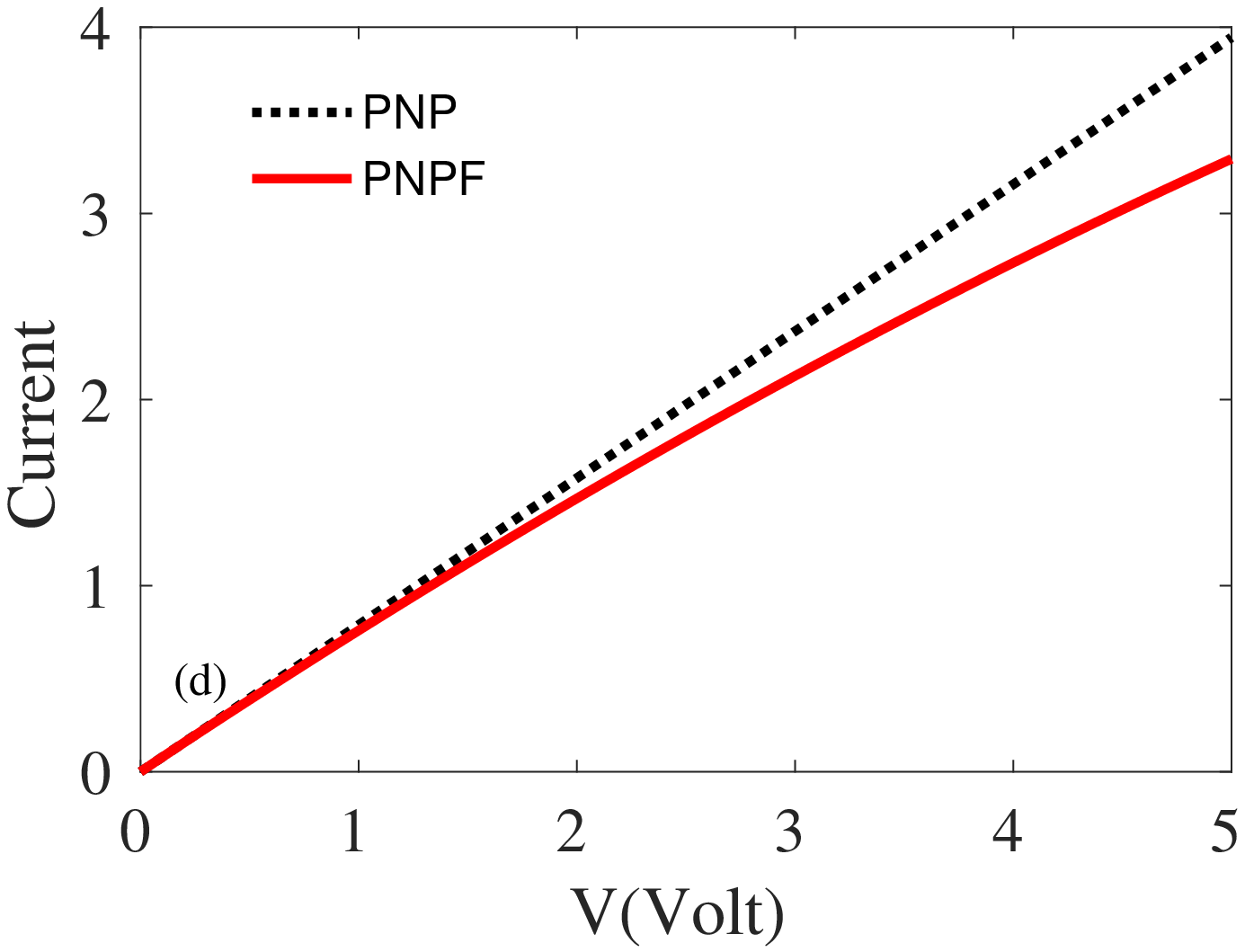}
\caption{When the system approaches to steady state. (a) $Na^+$ density distribution. (b) $Cl^-$ density distribution. (c) Temperature distribution. (d) Voltage-Current relation of the system. PNP stands for the classical Poisson--Nernst--Planck model, PNPF stands for the Poisson--Nernst--Planck--Fourier system with temperature effect.}
\end{center}
\end{figure}

The initial concentration and temperature distribution are all constants in space. Then we apply constant electrical voltage on the boundary, $\phi(0)=0$ and $\phi(L)=V$, so that the ions immigrate under the electrical field.  For the classical PNP system, where the temperature is a constant, we do not have the temperature equation. Under the same setup, the ionic densities in the classical PNP system remain to be homogeneous in space and time, thus the electrical potential becomes linear and ionic velocity is proportional to the electrical field. So the $VI$ curve of PNP is a linear function.

From panel (c) we can see that, the electrodiffusion can enhance the local temperature. With higher voltage applied, the entropy production as well as the Joule heating effect becomes more significant, so that we have higher temperature in steady state. And the ionic distributions in panel (a) and (b) are no longer homogeneous. The internal ionic density for both $Na$ and $Cl$ becomes lower than the Dirichlet boundary due to the thermal effect. As a consequence, although the diffusion efficiency increases with temperature, the overall ionic current reduces, as shown in panel (d).

\section{Conclusion} The temperature diffusion and mechanical diffusion are coupled for non-isothermal fluid. We have proposed a self-consistent framework to derive the equations for electrothermal diffusion, which can also be applied and generalized to many other systems with different kind of inter-molecular interaction. When applied to the imcompressible Navier--Stokes system, we obtain the Navier--Stokes--Fourier equations; when applied to the classical Poisson--Nernst--Planck system, we obtain the Poisson--Nernst--Planck--Fourier equations.  Our approach is consistent with laws of thermodynamics. The constitutive relation for the mechanical fluxes are governed by the force balance equations, where we use the pressure instead of the chemical potential. The thermal distribution is given by the heat equation with additional heat convection and heat sources.


\end{document}